\documentclass[aps, prl,longbibliography, superscriptaddress, preprint]{revtex4-1}

\usepackage{graphicx}

\setcitestyle{super}

\begin{document}
\title{Network analysis of synthesizable materials discovery}
\author{Muratahan Aykol}
\email[]{murat.aykol@tri.global}
\affiliation{Toyota Research Institute, Los Altos, CA 94022}
\author{Vinay I. Hegde}
\affiliation{Northwestern University, Evanston, IL 60208}
\author{Linda Hung}
\affiliation{Toyota Research Institute, Los Altos, CA 94022}
\author{Santosh Suram}
\affiliation{Toyota Research Institute, Los Altos, CA 94022}
\author{Patrick Herring}
\affiliation{Toyota Research Institute, Los Altos, CA 94022}
\author{Chris Wolverton}
\affiliation{Northwestern University, Evanston, IL 60208}
\author{Jens S. Hummelsh{\o}j}
\affiliation{Toyota Research Institute, Los Altos, CA 94022}
\date{\today}


\begin{abstract}
Assessing the synthesizability of inorganic materials is a grand challenge for accelerating their discovery using computations. Synthesis of a material is a complex process that depends not only on its thermodynamic stability with respect to others, but also on factors from kinetics, to advances in synthesis techniques, to the availability of precursors. This complexity makes the development of a general theory or first-principles approach to synthesizability currently impractical. Here we show how an alternative pathway to predicting synthesizability emerges from the dynamics of the materials stability network: a scale-free network constructed by combining the convex free-energy surface of inorganic materials computed by high-throughput density functional theory and their experimental discovery timelines extracted from citations. The time-evolution of the underlying network properties allows us to use machine-learning to predict the likelihood that hypothetical, computer-generated materials will be amenable to successful experimental synthesis.
\end{abstract}

\maketitle

\section{Introduction}
\begin{figure}
\includegraphics[width=0.95\linewidth]{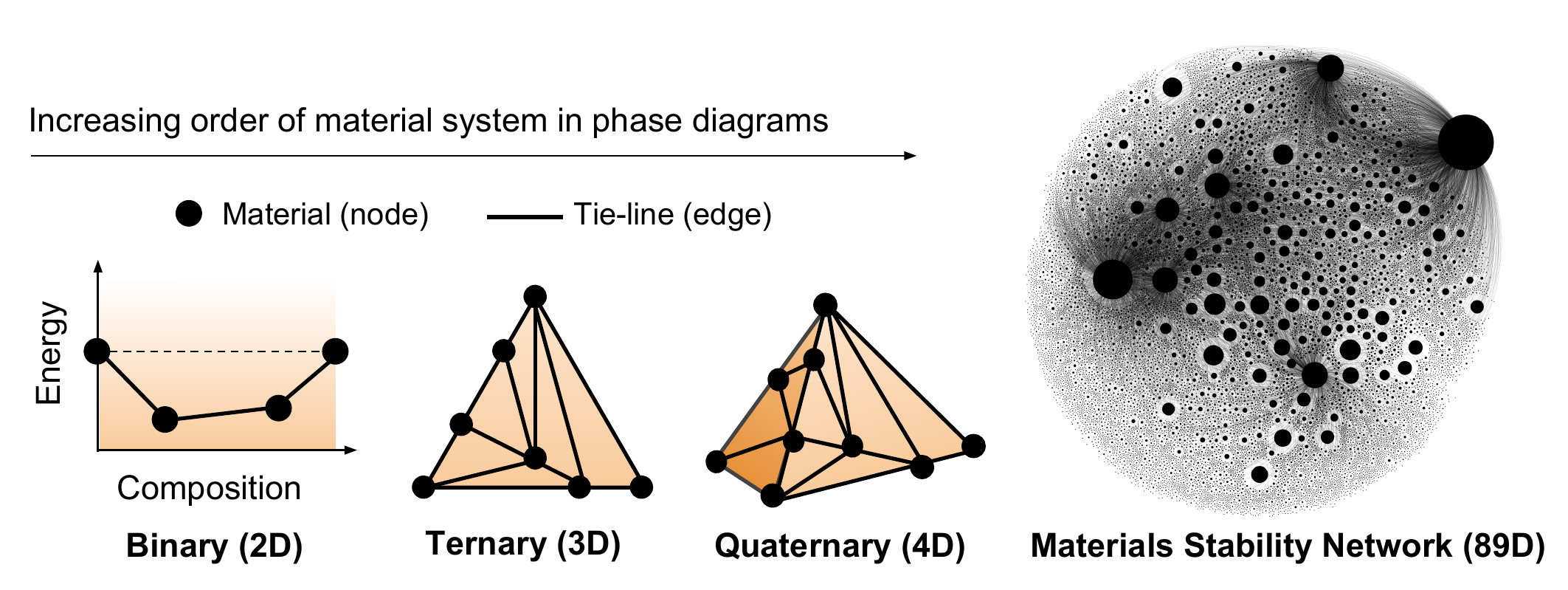}
\caption{\label{fig_synth} \textbf{Network representation of material phase diagrams.} The schematic illustrates phase diagrams with the order of the system ranging from 2-dimensional binary to the 89-dimensional materials stability network central to this work. The energy-composition convex-hull is shown for the binary system, and all higher order phase diagrams are projections of their respective $N$-dimensional convex-hulls to two dimensions, where materials are represented as nodes and tie-lines as edges. For clarity, only those tie-lines connected to high-degree nodes are shown in the materials stability network, where the sizes of the nodes are also scaled to reflect their degree.}
\end{figure}

Synthesis prediction for inorganic materials remains one of the major challenges in accelerating materials discovery,\cite{hemminger2015challenges,Sun2016,Kim2017,Aykol2018} mostly because the complexity of the synthesis process itself hinders the development of a general, first-principles approach to it.\cite{Kim2017, Alberi2018} Thermodynamic stability is one of the main factors that strongly influence synthesizability of a material, but extracting it requires the knowledge of energetics of competing phases. This bottleneck has recently been addressed for inorganic materials by high-throughput (HT) density functional theory (DFT) databases,\cite{Kirklin_unpub,Saal2013a,Jain2013,Curtarolo2012} which provide access to systematic DFT calculations of thousands of existing inorganic materials as well as hypothetical ones. These databases allow the construction of a comprehensive energy convex-hull: the multidimensional surface formed by the lowest energy combination of all phases. Phases that are on the convex-hull are thermodynamically stable, and tie-lines connecting two phases indicate two-phase equilibria. Given that it is composed of stable materials (nodes) connected by tie-lines (edges), the convex-hull is a naturally occurring thermodynamic network (Fig.\ref{fig_synth}), analogous to the world-wide-web, social, citation, and protein networks.\cite{Barabasi2001,Newman2001,Albert2002,Barabasi2009,Rzhetsky2015} The information encoded in this new network of materials can be harnessed with the tools provided by the emerging paradigm of network science, and form the basis of new data-driven models for outstanding materials challenges, such as predicting synthesizability.

The chronology of discoveries can reveal the dynamics of this network of materials. The discovery of a material can be associated with the physical identification and recording of a new crystal structure and chemistry for a target application or general scientific exploration. With this definition, to be traceable as discovered, a material should (i) physically exist, i.e. be amenable to synthesis or occur in nature, and (ii) have a record of structural characterization that can serve as a footprint for the onset of scientific interest. Both of these criteria can be traced from crystallographic databases,\cite{Belsky2002,Grazulis2009} which are dominated by structures of existing materials resolved with diffraction experiments. Assuming the time lag between the actual synthesis and/or characterization and the publication is not significant, the time of discovery of a material, and in particular the implied \textit{successful synthesis}, can be approximated to be the earliest cited reference available in such collections (See Methods).

The thermodynamic information encoded in the convex-hull is important but not sufficient to explain the successful synthesis and discovery of a material.\cite{Aykol2018} On the other hand, the collective influence of all complex factors on synthesizability is already reflected in the measured ground truth: whether a material was synthesized or not. Thus, when combined with the historical records on the time of discovery of existing materials, the dynamics of the resulting temporal stability network encodes also the \textit{circumstantial} information beyond thermodynamics that influences discovery. Such information implicitly includes scientific and non-scientific effects almost impossible to capture otherwise at this scale, such as the availability of kinetically favorable pathways, development of new synthesis techniques, availability of new precursors, changes in interest or experience of researchers in a particular chemistry, structure or application, and even changes in policies that influence research directions. Here we combine the stability information from HT-DFT with the citation-extracted discovery timeline, both available in the Open Quantum Materials Database (OQMD),\cite{Saal2013a, Kirklin_unpub} and determine the temporal evolution of the stability network as more materials are discovered and added to it. Using the extracted network properties of materials, we demonstrate how a model can be developed to estimate the likelihood of synthesis of new, computationally-predicted stable materials.

\section{Results}

\subsection{The materials stability network and its time evolution}

\begin{figure}
\includegraphics[width=0.7\linewidth]{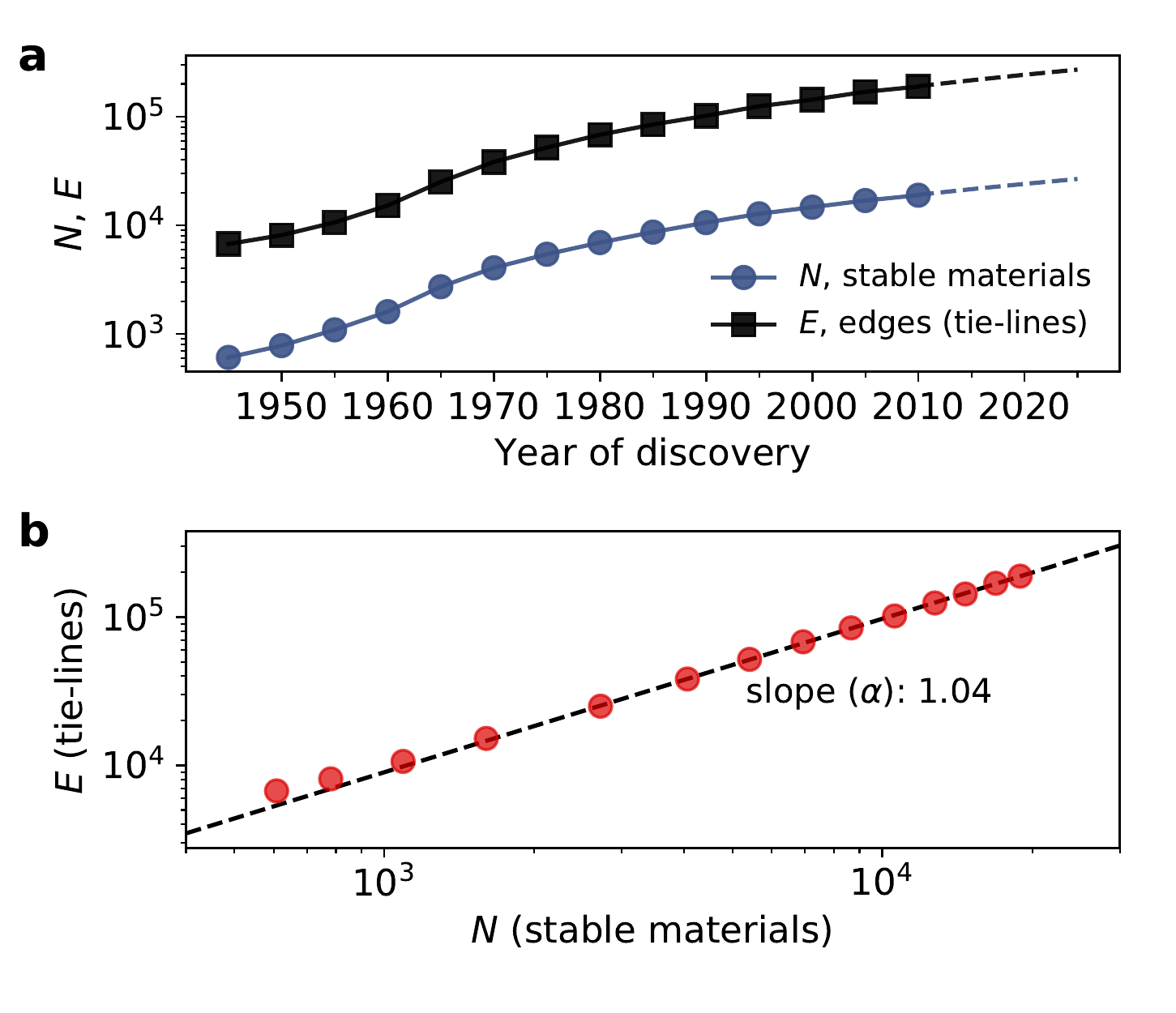}
\caption{\label{fig1} \textbf{Evolution of the size of the materials stability network.} (a) Time evolution of the number of stable materials (i.e. nodes), $N$, and tie-lines (i.e. edges), $E$, and (b) how the number of nodes and tie-lines vary with respect to each other. A tie-line is included in the evolving network only after both nodes it is connecting are identified as discovered. Dashed-lines in (a) are extrapolations of $N$ and $E$ from the available data (markers and solid-lines) by fitted quadratic polynomials. Dashed-line in (b) is a linear fit to the data (circles). Fits performed in both panels exclude the first four times steps to obtain fits that are more representative of more recent times. A plot of number of stable materials discovered each year as a function of time is also available in Supplementary Fig. 1.}
\end{figure}

The complete network formed by the current convex-hull in the chemical space of all elements is extremely dense with 41 million tie-lines.\cite{Hegde2018} To find the most relevant set of tie-lines for synthesis, we subsample this network to obtain those that control the stability of at least one material, i.e. those in chemical subspaces where there is at least one stable material inside the composition simplex (See Methods). This process yields an informative subset of $\sim2\times10^{5}$ tie-lines for synthesis that is also computationally tractable for repeated analysis, essential for building a predictive model as described later. Hereafter, we refer to this subset as the materials stability network to differentiate it from the complete network. We then trace retrospectively how this network was uncovered over time until it reached its present state. The number of stable materials discovered, $N$, and the number of tie-lines defining their equilibria as described above, $E$, are both growing with time (Fig.\ref{fig1}a and Supplementary Fig. 1). A polynomial fit to $N(t)$ shows the number of stable materials discovered by year 2025 will reach $\sim27\times10^{3}$ from the present number of $\sim22\times10^{3}$. The rate of stable materials discovery is $\sim$400 year$^{-1}$ today and projected to reach $\sim$540 year$^{-1}$ by 2025, suggesting that the discovery of stable materials is accelerating. $E$ is increasing faster than $N$ (Fig.\ref{fig1}b), with $\alpha\approx1.04$ in the densification power-law $E(t)\sim N(t)^{\alpha}$.\cite{Leskovec2007} Thus, the materials stability network is getting denser, which may be explained by researchers discovering materials closely connected with those already known, using the latter as stepping stones for the synthesis of new ones,\cite{Leskovec2007,Rzhetsky2015} while uncovering the underlying ultimate network.\cite{Pedarsani2008}

\begin{figure}
\includegraphics[width=0.7\linewidth]{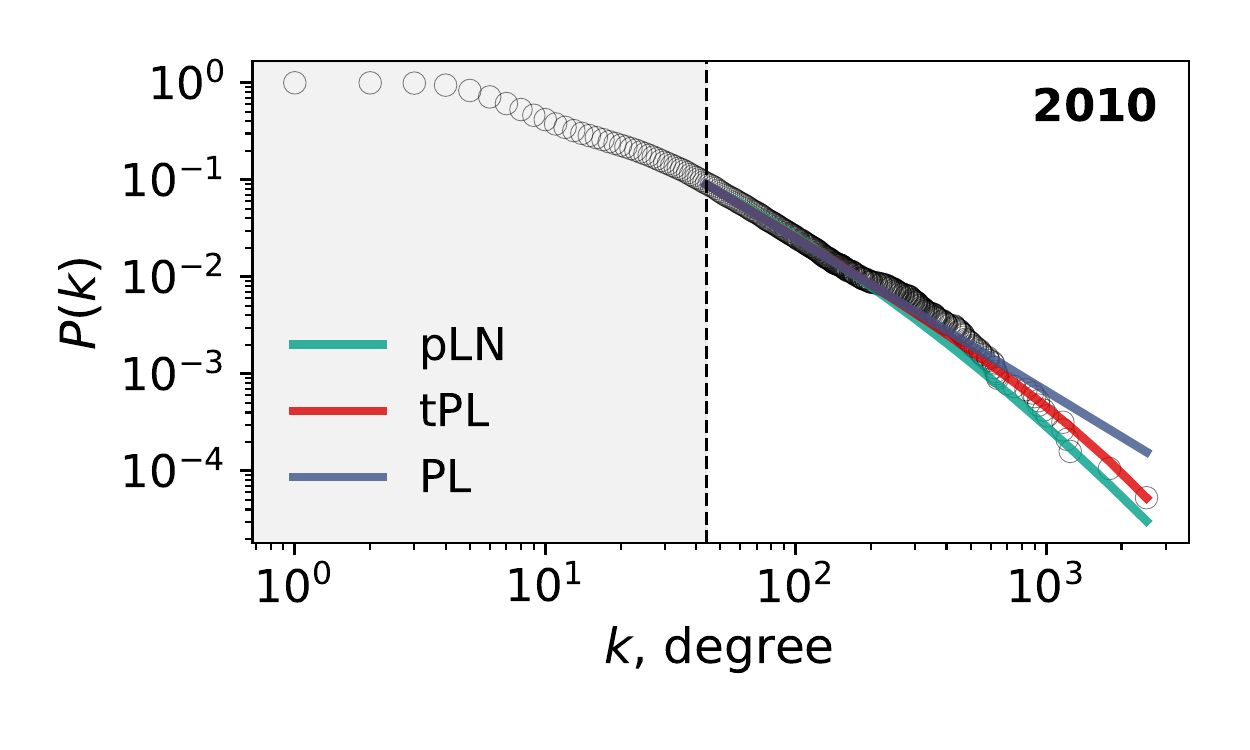}
\caption{\label{fig2} \textbf{Degree distribution among stable materials discovered by the year 2010.} The complementary cumulative distribution function ($P(k)$) of the degree distribution $p(k)$ of stable materials (circles) is plotted along with the fitted distributions (solid lines). Each point $P(k)$ represents the probability that a material has greater than $k$ tie-lines connected to it in the network. Power-law, truncated power-law (with exponential cutoff) and positive log-normal distributions are labeled as PL, tPL and pLN, respectively. The dashed line shows $k_{min}$, the lowest degree used in fitting. Degree distributions of other times are shown in Supplementary Fig. 1.}
\end{figure}

The degree distribution, $p(k)$, where $k$ is the degree of each node, is one measure of the topology of networks. Here $k$ corresponds to the number of tie-lines a material has. In recent years, scale-free networks that obey a power-law distribution, $p(k)\sim k^{-\gamma}$, have received significant attention.\cite{Barabasi2009} While the materials stability network is far from a power-law in early times (e.g. 1960s), it has evolved into a distribution close to it, as shown in Fig.\ref{fig2} for 2010 (Supplementary Tables 1-2 and Supplementary Fig. 2). The exponent $\gamma$ becomes constant at $2.6\pm0.1$ after the 1980s (Supplementary Fig. 3), within the range $2<\gamma<3$ as the other scale-free networks like the world-wide-web or collaborations\cite{Albert2002,Barabasi}.

This scale-free character hints at the presence of hubs with significantly larger $k$ compared to other nodes and a robust network connectivity,\cite{Albert2000} implying that materials missing randomly from the network (because they have not been discovered yet) are not expected to hinder the discovery of others. However, if there are missing hubs,\cite{Barabasi2009} new material classes disconnected from the present network may be awaiting discovery. The biggest hub here is O$_2$ with nearly 2,600 tie-lines, followed by Cu, H$\mathrm{_2}$O, H$_2$, C and Ge with more than 1,100 tie-lines. Elemental N$_2$, Ag, Si, Fe, Se, Mn, Co, K, Te, and Bi and oxides BaO, CaO, Li$_2$O, SrO, Cu$_2$O, MgO, SiO$_2$, La$_2$O$_3$, Al$_2$O$_3$, CuO, MnO, ZnO, Y$_2$O$_3$, Nd$_2$O$_3$, Mn$_3$O$_4$, Sc$_2$O$_3$, Gd$_2$O$_3$, Mn$_2$O$_3$, FeO, Fe$_2$O$_3$, Cr$_2$O$_3$, NiO, BeO, V$_2$O$_3$, and VO$_2$ are densely connected with 350 or more tie-lines. These are the species that play a dominant role in determining stabilities, and subsequently influencing synthesis of many materials, whether as starting materials, decomposition products of precursors, or simply as competing phases. 

Analysis of the discovery timelines indicate that the number of new stable oxygen-bearing materials has been increasing exponentially and faster than all other chemistries, which is in line with the observed predominance of oxides as hubs, and also correlated with the historically high average-degree of O-bearing materials (Supplementary Fig. 4). These trends follow the widely-accepted Barabasi-Albert model for the growth of scale-free networks,\cite{Barabasi, Barabasi1999} where a small difference in the node degrees in the early days gets drastically amplified over time, because of the preferential attachment of new nodes to higher-degree nodes. These results also indicate that identifying new hubs in chemistries such as pnictides, chalcogenides, halides or carbides may accelerate discovery in those spaces. To corroborate this hypothesis further, we compared several such chemistries with oxygen (Supplementary Fig. 4) and observed that as more materials with hub-like character emerged among the phosphorus-bearing materials in 1960s (as reflected in their average degree), the discovery in this space accelerated, with a notable upsurge in number of new P-bearing stable materials in later years.

\subsection{Prediction of materials discovery from network dynamics}

\begin{figure*}
\includegraphics[width=\linewidth]{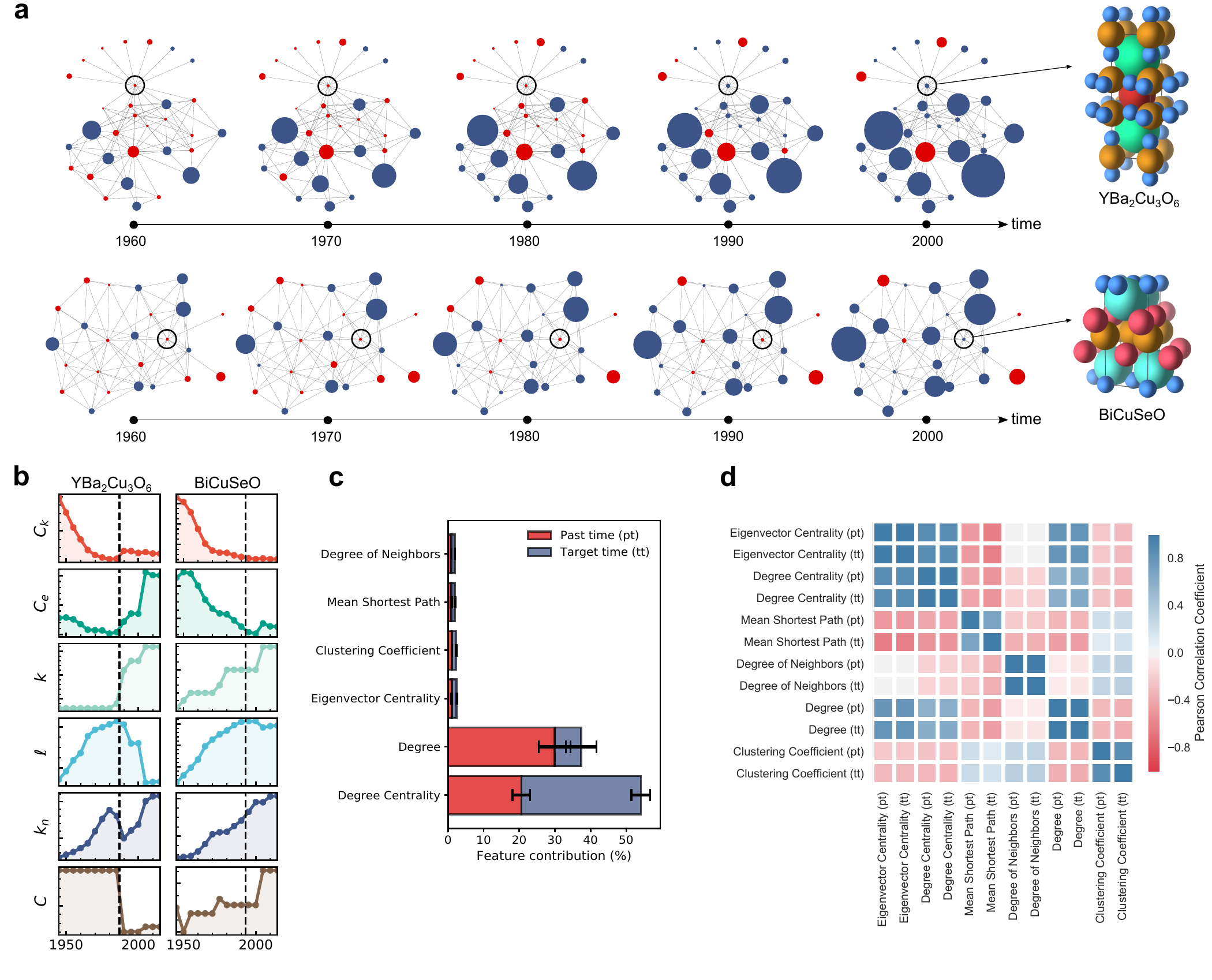}
\caption{\label{roc} \textbf{Network evolution and properties of the machine-learned synthesizability models.} (a) Time evolution of the local environments of two sample materials (marked with open circles), superconductor YBa$_2$Cu$_3$O$_6$ and thermoelectric BiCuSeO, in the materials stability network. Materials (nodes) discovered by a given temporal state of the network are shown in blue, whereas those awaiting discovery are red. Node size is proportional to degree. (b) Time evolution of the network properties of sample materials YBa$_2$Cu$_3$O$_6$ and BiCuSeO; namely, degree and eigenvector centralities ($C_k$ and $C_e$), degree ($k$), mean-shortest-path ($\ell$), mean degree of neighbors ($k_n$) and clustering coefficient ($C$), where the vertical dashed lines show the approximate time of discovery. (c) Feature contributions to the RF model as derived from the Gini importance. (d) Pearson correlation coefficients of time-dependent network properties used in models as features, where pt and tt denote past time and target time, respectively, corresponding to a given sequence of window size of two (See Methods). Variables and names of network properties are used interchangeably in (b), (c) and (d).}
\end{figure*}

While the evolution of global properties of the network is slow (Supplementary Fig. 5), the network properties of individual nodes are evolving rapidly as their local-environments change, as exemplified in Fig.\ref{roc}a-b for high-temperature superconductor YBa$_2$Cu$_3$O$_6$ and high-ZT thermoelectric BiCuSeO. Since this temporal evolution encodes circumstantial factors beyond thermodynamics that may contribute to discovery (and synthesis), properties that characterize the state of a material in relation to the rest of the network can realize a connection between these explicit or implicit factors and its discovery.

To reproduce that connection, we turn to designing a machine-learning model based on the network properties of materials, which we will then use to predict likelihood of synthesis of hypothetical materials: those created on the computer but have never been made. The present stability network has about 22,600 materials, of which $\sim$19,200 are physically-existing materials from crystallography databases and can be used in model building, and $\sim$3,400 are hypothetical, generated via high-throughput prototyping.\cite{Kirklin_unpub, Kim2017a, Emery2016, Ma2017} Prediction of synthesis likelihoods in the latter category can help bridge the gap between computational discovery and the real world.

We use six network properties for each material in model training; namely, degree and eigenvector centralities, degree, mean shortest path length, mean degree of neighbors, and clustering coefficient (Fig. \ref{roc}b and Supplementary Fig. 6). Degree and eigenvector centralities reflect the relative importance of a node in influencing stabilities, emphasizing the number of connections and importance of neighbors, respectively. We normalize these metrics such that they are mostly independent of the size of the network.\cite{Ruhnau2000} We find that degree without normalization is also useful for capturing the influence of the temporal state of the network on connectivity. Mean shortest path length, the mean of minimum number of tie-lines to travel from a node to every other node, and mean degree of neighbors serve as a proxy for ease of access to a particular material in synthesis. The clustering coefficient indicates how tightly-connected the neighborhood of a material is and may capture the local environment more immediately related to its synthesis.

\begin{figure}
\includegraphics[width=0.65\linewidth]{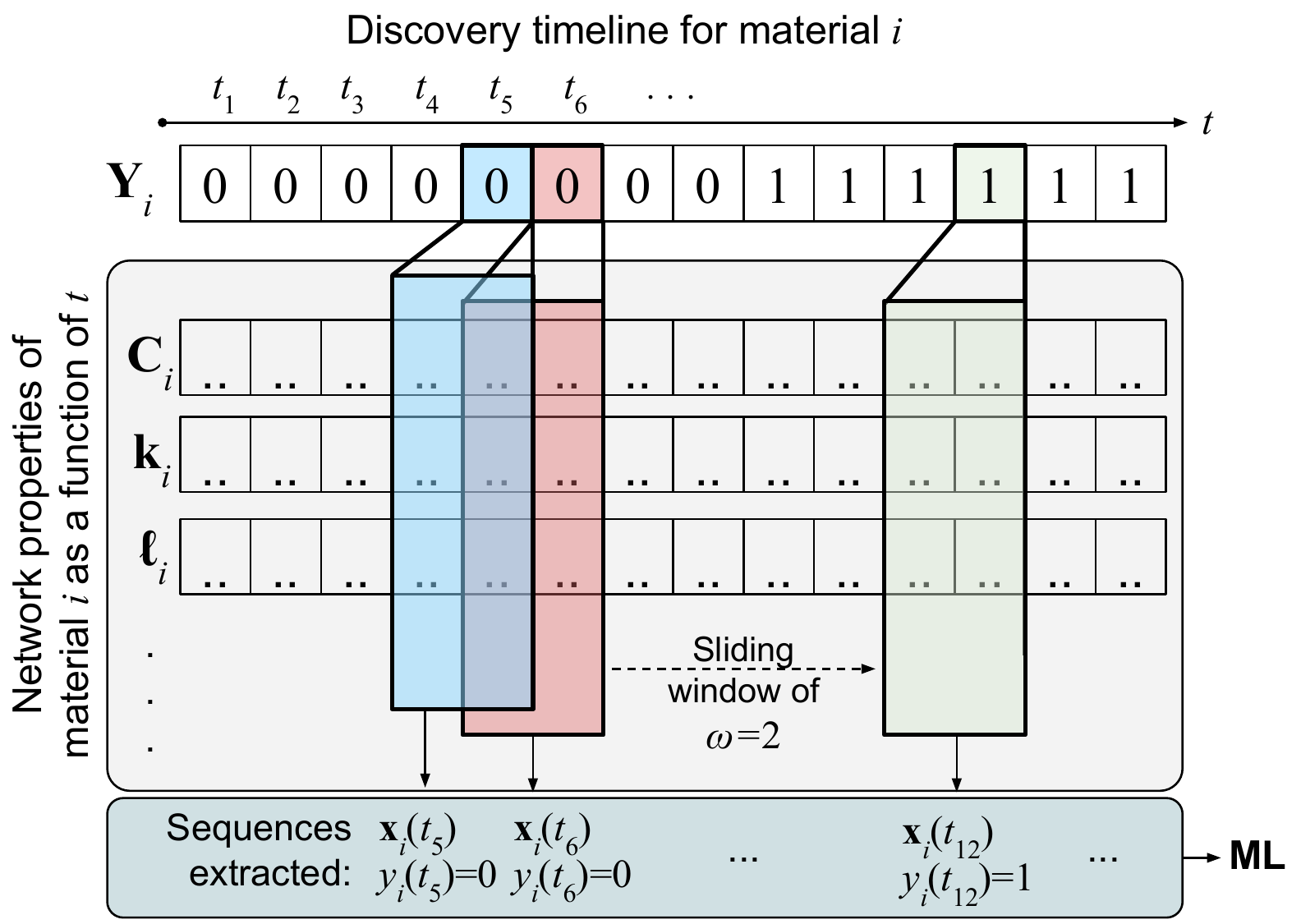}
\caption{\label{fig_slide} \textbf{Extraction of sequences from temporal network property data using a sliding window to use as input for machine-learning.} The vector $\mathbf{Y}_i$ stores the targets to be learned for material $i$, \textit{i.e.} encoding whether $i$ is discovered by a given time-step $t$ or not (as binary labels 1 and 0). $C_i$, $k_i$ and $\ell_i$ are examples for vectors of different network properties, encoding how those properties change over time as the network evolves, as explained in the text. The process of applying a sliding window (here with a width of $w=2$) to extract sequences of features and targets ($\mathbf{x}_{i,t}$, $y_{i,t}$) is illustrated. ML stands for the machine-learning task of training and testing classification algorithms using the extracted data.}
\end{figure}

Discovery is a time-irreversible event and its prediction is not a standard machine-learning problem in materials science. Here we use time-evolution of the aforementioned network properties of materials as features to form the basis of a sequential supervised-learning problem.\cite{Dietterich} We adopt a sliding-window approach to train experimental discovery classifiers and estimate likelihood of discovery, and the implied synthesis for synthetic materials (See Fig.\ref{fig_slide} and Methods). We employ two classification algorithms: $L_2$-norm regularized logistic regression (LR) for well-calibrated probabilities, and random forest (RF) for classification accuracy. The subclass of sequences where a material changes from undiscovered to discovered in the time domain represents the rare-event of discovery in sequential data, for which we define stringent event-detection metrics for precision and recall using a prediction period approach (See Methods),\cite{Weiss1998} and found these metrics for LR and RF to be near 30\% and 90\%, respectively, for detection within $\pm$1 time-step and close to 50\% and 95\% for detection within $\pm$2 time-steps (where larger prediction periods make the correct classification easier), outperforming baseline detection metrics by a significant margin (Supplementary Fig. 7). As another performance evaluation for the present problem, distributions of the difference between the estimated time-step of discovery (See Methods for how this estimation is done from classification results) and its true value ($\Delta t$) for the two models tested in the present work are compared in Supplementary Fig. 8, along with baseline models. For both LR and RF models, distributions are centered close to zero, with LR showing a tendency to estimate slightly earlier times with a mean $\Delta t \approx -1.6$ whereas RF delivers more precise estimates with mean $\Delta t \approx -0.2$ (each time-step is 5-years long in the present work). Baseline models yield distributions with means far from zero. LR has a much broader distribution than RF, however, with a standard deviation at $\sim$3.5 time-steps, whereas that of RF is at $\sim$1.2 time-steps.

To understand how these models provide accurate predictions for synthesizability, we investigate the correlations between the network properties, and how much they contribute to predictions (Figs.\ref{roc}c-d). Except the eigenvector centrality with degree or degree centrality, distinct features are not too strongly correlated. Identical features within a time sequence are naturally more correlated (e.g. degrees in a sequence), but distinct enough for the models to utilize them (Supplementary Fig. 9). Confirming the significance of tie-lines in influencing synthesis, degree and degree centrality, two closely connected but not highly-correlated metrics, play the biggest roles in decision-making, with substantial contributions adding up to $\sim$90\%. The rest of the features still play a non-negligible role, providing the remaining 10\%.

\section{Discussion}

The trained models can be used in multiple ways; for example, to predict class labels or probabilities for synthesizability in network environments pertaining to the present time or a past time. For the present time, models predict about 93\% of hypothetical materials in the network have a synthesis probability, $p$, greater than 0.5. This prediction is in line with the notion that stable materials in HT-DFT databases are likely to be more amenable to synthesis. However, synthesis is a costly process and knowing its likelihood of success is critical before an attempt in the laboratory. Using LR's calibrated probabilities with RF's more accurate classifications within the intersection of positive classification sets they predict (92\% identical), we find that out of $\sim$3,400 stable hypothetical materials present, only about 10\% have $p>0.95$ for immediate synthesis (Supplementary Note 1).

Our approach can assist the decisions on where to allocate synthesis resources after computational design. For example, Kim et al.\cite{Kim2017a} performed a computational search and suggested new high-capacity Li$_4$ABO$_6$ cathodes for Li-ion batteries (A and B represent different elements), for which we find the likelihoods of synthesis to range from $p=0.52$ for Li$_4$SbRhO$_6$ to $p=0.85$ for Li$_4$NiTeO$_6$. In fact, several of those predictions with $p>0.6$ were synthesized.\cite{Bhardwaj2014,Sathiya2013} For the novel ABO$_3$ perovskites identified in two recent computational studies,\cite{Emery2016,Balachandran2018} we find the synthesis likelihoods to range from $p=0.54$ for Pu$_{}$Ga$_{}$O$_{3}$ to $p\approx1$ for EuGeO$_3$. For several of such predictions with $p>0.9$, reports of synthesis exist.\cite{Shukla2010,Akamatsu2012} For the inverse-Heusler alloys uncovered in a HT search for spintronics,\cite{Ma2017} we predict $p$ to be in the range 0.56 (Ti$_{}$In$_{}$Co$_{2}$) - 0.94 (Fe$_{}$Ge$_{}$Ru$_{2}$) (a complete list of probabilities is available in Supplementary Table 3). Today such computational studies can rapidly identify hundreds of new hypothetical materials with target functionalities, but the cost and complexity of synthesis often hinders systematic attempts for their realization. The ability to predict synthesis likelihoods is expected to bridge this gap between computational and experimental research groups.

The network-based models can also be used to invert the discovery predictions and find at what point in time a hypothetical material could have already been made. Based on the RF model, we estimate that only about 10\% of the stable hypothetical materials that are predicted to be synthesizable today were synthesizable by 2005, and only about 30\% were synthesizable by 2010 (Supplementary Fig. 10), implying the progress within the last 10--15 year period have improved their chances of synthesis.

Similarly, since most of the materials discovered in the last few decades have likely been made with contemporary synthesis methods developed or improved since the mid-20$^\mathrm{th}$ century (from sol-gel to advanced deposition methods to new precursors), models trained only with earlier discoveries should intuitively predict a majority of the newer materials as unlikely to be made in the distant past. Indeed, a model trained only with discoveries up until and including year 2000 yields probability distributions that consistently shift to higher values with time for materials discovered after year 2000 (unseen to model) as shown in Supplementary Fig. 11, confirming that the predictions agree with our intuition. Besides, almost all materials that were discovered after 2000 are predicted to have $p>0.4$ in year 2000 with this model; i.e. materials with $p<0.4$ in 2000 were rarely synthesized after 2000. However, the application of the models to new materials assume that the mechanisms of materials discovery continue to follow similar trends to those in the past and present, and therefore by design the models cannot forecast the future. For instance, we observe that the probability distributions predicted by the above model trained with material discoveries until 2000 cannot clearly differentiate between progression of most materials discovered in its future (except a fraction of materials near $p\approx1$, where the predictions look correlated with the discovery timeline), confirming future timeline forecasts cannot be made for most of the materials (Supplementary Fig. 11). Ultimately the decisions on which materials to make are made by the scientists and the future is shaped accordingly. The models merely provide statistical predictions based on the latest network data they are exposed to, within the limits of their underlying approximations. 

Given the advances in materials discovery techniques including complex and high-throughput experimental or simulation capabilities, intuitively, the present models are likely a lower bound for the future synthesis likelihoods, as long as the non-scientific factors, such as the science policies and funding remain sustainable. Demonstrating how network science and machine learning can be combined to build predictive methods for materials, we expect the present work to pave the way for new, improved methods for materials discovery, possibly addressing synthesizability in the unbounded space of metastable materials (which would require constructing linkage rules beyond the convex-hull), or examining applications beyond synthesis.

\section*{Methods}
\subsection{Network data and analysis}
The network presented in this work is constructed from the energy-composition convex-hull of OQMD, which is a collection of systematic DFT calculations of inorganic crystalline materials, and subsequent properties derived from them, such as formation energies.\cite{Saal2013a, Kirklin_unpub} DFT is known to provide a good compromise between accuracy, especially in terms of determining relative stabilities of materials, and computational cost, and is the current state-of-the-art for first-principles high-throughput computations of materials.\cite{Hautier2012, Lejaeghere2016, Aykol2018} We used the version 1.1 of the OQMD data available at http://oqmd.org.

The NetworkX package was used for the calculation of the network properties.\cite{Hagberg2008} The maximum-likelihood method was used to fit the distributions and the $k_{min}$ values were found by minimizing the Kolmogorov-Smirnov distance.\cite{Clauset2009,Alstott2014} The \textit{powerlaw} library was used in fitting the distributions.\cite{Alstott2014} Goodness-of-fit comparisons of different distributions are available in Supplementary Table 1. The method for subsampling of the complete network to obtain the materials stability network is further described in Supplementary Methods.

\subsection{Model construction}
To prepare the input vectors for training the machine learning models, we create multiple sequential training examples $(\mathbf{x}_{i,t}, y_{i,t})$ for each material $i$ from its temporal data, where feature vector for time $t$, $\mathbf{x}_{i,t}$ extends to features for the past times within a window $w$, and where the target $y_{i,t}$ encodes binary labels 1 and 0 respectively indicating whether a material was discovered at that point in time or not (Fig.\ref{fig_slide}). We adopted $w=2$ for the present work. We analyzed the networks with 5-year increments starting from 1945, and found that a window of width $w=2$ (\textit{i.e.} encompassing 10 years) provides sufficient prediction accuracy, without any need for recurrence (i.e. including past $y$ as part of $\mathbf{x}$). Network properties of a material pertaining to times when it was undiscovered are calculated by hypothetical, individual insertion of that material into the materials stability network (as if it existed at that point in time). Further details of each step in model creation can be found in Supplementary Methods. Discovery times of known materials are approximated by their earliest dated reference for their structures reported in the ICSD,\cite{Belsky2002} except for the elemental references, which are defined as discovered ($y=1$) at all times. We expect this approximation to (i) reasonably hold given the coarse-enough discretization of the timeline (5-years) would already account for the typical delays between characterization and publication (e.g. 1--2 years), and (ii) not significantly affect the model training as the delays are likely to be in the form of a nearly constant shift for all materials, as one might intuitively expect the distribution of the delays to be narrow and centered around 1--2 years at most.

\subsection{Model training and evaluation}
Model training and parts of the evaluation were performed using scikit-learn.\cite{scikit-learn} In training the LR models, we use a larger weight (approximately 2.5 times) for the minority subclass of $y=[0,1]$ (i.e. a material transitioning from undiscovered to discovered), compared to the other subclasses to obtain evenly distributed accuracies across all subclasses. RF models use 200 estimators. Models were also tested against baseline classifiers including class distribution prediction, majority and/or constant class prediction and random classifiers, and found to outperform all. Feature importance in RF model is calculated as the Gini importance. Calibration is applied to model probabilities; however, given the approximate nature of this process and the variability in absolute values of the resulting probabilities, probabilities should be considered mostly to reflect relative likelihoods among materials. We use 5-fold cross-validation (CV) for the evaluation of all models.

We follow event-based strategies for model evaluation that consider the entire timeline of the materials and test/train splitting of sequences $(\mathbf{x}_{i,t}, y_{i,t})$ (where $i$ is a material and $t$ is the time-step) is accordingly performed at the material-level. The standard model evaluation metrics in classification are defined as: $\mathrm{precision = TP/(TP+FP)}$, $\mathrm{recall = TP/(TP+FN)}$, F1-score $= 2 \times \mathrm{precision}\times \mathrm{recall}/(\mathrm{precision}+\mathrm{recall})$, where TP, FP and FN are the number of true positives, false positives and false negatives, respectively. While these metrics help evaluate the performance of algorithms in classifying targets as 0 or 1 in a standard way (and are all above 90\% and 70\% respectively for RF and LR models in material-level splitting and CV), for the detection of the discovery itself (\textit{i.e.} the transition from 0 to 1), which can be described as a rare-event detection in sequential data, modified definitions for precision and/or recall are more suitable.\cite{Weiss1998} In the present event-based strategy, the model is assumed to predict a discovery event at the very first positive prediction it makes in the timeline of that material, and metrics like TP and FP count whether the target event is captured at the correct time or not, and precision is calculated accordingly with the same formula as above. Recall becomes the fraction of target events correctly captured: TP/(Total-number-of-discovery-events). These metrics, however, would equally weight FP's made one step away from the ground-truth, vs., for example, five steps away from it, where the latter case can be considered worse from a practical point of view. To partially address this issue, we employ a prediction-period concept,\cite{Weiss1998} where a $\pm$ time-range is defined such that a discovery prediction would be considered TP if the correct discovery time falls in that range, or FP if it falls outside. These metrics then become a function of the size of the prediction period as we show in Supplementary Fig. 7. Another approach for model evaluation is the direct comparison of the difference between the predicted discovery times and their actual values as shown in Supplementary Fig. 11, where again the model is assumed to make the discovery prediction at the very first positive classification it predicts for a material.

\section*{Acknowledgements}
V.I.H. acknowledges support from Toyota Research Institute through the Accelerated Materials Design and Discovery program. C.W. acknowledges the support of the National Science Foundation, through the MRSEC program, grant number DMR-1720139.

\section*{Author contributions}
M.A. conceived and designed the project, with input from J.S.H., S.S., P.H. and L.H. M.A. performed the network analysis and machine-learning with input from V.I.H, C.W., S.S., P.H. and L.H. M.A., V.I.H. and C.W. computed the convex-hull. All authors contributed to the discussions of the results. M.A. wrote the manuscript with input from all authors.

\section*{Competing interests}
M.A., L.H., S.S. and P.H. filed a patent application on network-based synthesis prediction: US App. No. 16/004,232 on 08 June 2018.

\section*{Data availability}
Data used in this work is provided as Supporting Data and can also be accessed at https://data.matr.io/2.

\end{document}